\documentclass[a4paper,fleqn,usenatbib]{mnras}

\usepackage[T1]{fontenc}
\usepackage{ae,aecompl}

\usepackage{graphicx}	
\usepackage{amsmath}	
\usepackage{amssymb}	

\def\msun{{\rm\,M_\odot}}

\def\gtsima{$\; \buildrel > \over \sim \;$}
\def\simgt{\lower.5ex\hbox{\gtsima}}

\def\msun{\hbox{M$_\odot$}}
\def\reference{In: Pei, J., Tseng, V.S., Cao, L., Motoda, H., Xu, G. (eds) Advances in Knowledge
Discovery and Data Mining. Lecture Notes in Computer Science(), vol 7819. Springer, Berlin,
Heidelberg.}

\title[A disrupting open cluster]{Catching a Milky Way open cluster in its last breath}

\author[Andr\'es E. Piatti]{
Andr\'es E. Piatti$^{1,2}$\thanks{E-mail: andres.piatti@unc.edu.ar} \\
$^{1}$Instituto Interdisciplinario de Ciencias B\'asicas (ICB), CONICET-UNCUYO, 
Padre J. Contreras 1300, M5502JMA, Mendoza, Argentina\\
$^{2}$Consejo Nacional de Investigaciones Cient\'{\i}ficas y T\'ecnicas, Godoy Cruz 
2290, C1425FQB,  Buenos Aires, Argentina\\
}

\date{Accepted XXX. Received YYY; in original form ZZZ}

\pubyear{2022}

\begin{document}
\label{firstpage}
\pagerange{\pageref{firstpage}--\pageref{lastpage}}
\maketitle

\begin{abstract}
Theoretical models have suggested peculiar velocity dispersion profiles of
star clusters facing dissolution. They predicted that, besides bound stars
that still belong to the star cluster, and unbound ones already stripped off, 
there is an intermediate population of stars that
having acquired the enough energy to escape the cluster are still within the
cluster Jacobi radius. Both, potential escapers and unbound stars, show
hot kinematics, not observed along tidal tails of star clusters. We
report on the first evidence of an open cluster with stars crossing such
a transitional scenario, namely: ASCC~92. The open cluster gathers
nearly 10 percent of its initial total mass, and is moving toward
Galactic regions affected by higher interstellar absorption.
Precisely, the obscured appearance of the cluster could have hampered
disentangling its true internal dynamical evolutionary stage, previously.
\end{abstract} 




\section{Introduction}

 The All-Sky Compiled Catalogue of more than 2.5 million stars \citep[ASCC-2.5][]{k01} was 
used by \citet{kharchenkoetal2005} to identify known open clusters. They developed
an iterative pipeline to be used in that search, which they took advantage to discover other 109 new
open clusters. We focus here on ASCC-92, one of those new open clusters, that was also 
discovered from an analysis of the Tycho 2 catalogue \citep{hog2000} and the Digitized Sky Survey 
(DSS) plates, and named Alessi~31 \citep{alessietal2003,kronbergeretal2006}. The
cluster was also named MWSC~2723 by \citet{kharchenkoetal2012}. Surprisingly, the
cross-match of the cluster names was performed for the first time very recently by \citet{lx2019}.
Indeed, there have been several studies on the cluster without acknowledging those
that relies on different cluster names. 

ASCC-92 caught our attention while analyzing it from the {\it Gaia} DR3
\citep{gaiaetal2016,gaiaetal2022b} database, because some of the derived cluster's properties
hinted at, as far as we are aware, to be the first observed star cluster confirming the theoretical
predictions by \citet{kupperetal2010b}. \citet{kupperetal2010b} performed a comprehensive
set of N-body simulations to study the evolution of surface density and velocity dispersion 
profiles of star clusters as a function of time until cluster dissolution. They modelled clusters 
for an interval of 4 Gyr, if they did not dissolve before reaching this age. without stellar evolution 
and without primordial binaries. They found that potential escapers 
- stars energetically unbound located inside the Jacobi radius ($r_J$) - are more numerous than bound
stars at distances from the cluster centre $\ga$ 0.5$\times$$r_J$, and that beyond $\sim$ 
0.7$\times$$r_J$ they compose nearly the whole cluster population. They also showed from the
fit of nearly 104 computed surface density profiles that it is possible to derive reliable $r_J$  values
using King models for extended clusters on nearly circular orbits, and that by including to them three
more free parameters, it is possible to derive $r_J$ values with an accuracy of 10 per cent for
clusters in eccentric orbits. Likewise, they studied tidal debris in the cluster's outskirts and 
found that they are well represented by a power law with slope of -4 to -5, while close to 
apogalacticon it turns significantly shallower ($\le$ -1).

Numerical simulations of the 3D distribution of stars in clusters with tidal tails
have been previously developed, for instance, by \citet{kharchenkoetal2009b}. They have shown 
to provide a satisfactory representation of stars along the leading and trailing tails in the Hyades 
\citep{roseretal2019} and Preasepe \citep{rs2019}, respectively, uncovered using {\it Gaia} data. 
Likewise, {\texttt AMUSE} 
\citep[][and references therein]{pzm2018}, an astronomical multipurpose software environment
that simulates the evolution of a Hyades-like cluster on a relistic orbit, has
been recently used to confirm long tails in the Hyades \citep{jerabkovaetal2021} discovered from
{\it Gaia} data, while
the N-body code {\sc PETAR} \citep{wangetal2020a} combined with {\sc GALPY} \citep{bovy2015},
was employed by \citet{wj2021} to study the long-term evolution of young open clusters and
their tidal streams. As can be seen, the simulations by \citet{kupperetal2010b} stands out
in the sense that they predict a particular stage in the cluster internal dynamical evolution, 
rather than the existence of tidal tails.

In this work, we carried out a thorough analysis of ASCC~92 not easy handled open cluster. We
discovered that it is entering into an interstellar cloud and, more importantly, it is facing
its last breath before total disruption. As far as we are aware, there were not
studied open clusters cautch in such a final evolutionary stage. Moreover, the present results
provide the first evidence confirming theoretical speculations on the bound/unbound
conditions of cluster stars from the analysis of their total energies. In Section 2 we
describe our analysis and derive the cluster fundamental parameters, while in Section 3 we
discuss the obtained results to the light of previous studies on the cluster and uncover
the real disrupting cluster stage. Section 4 summarizes the main conclusions of 
this work.

\begin{table*}
\caption{Properties of ASCC-92 (Alessi~31).}
\label{tab1}
\begin{tabular}{@{}lcccccccc}\hline
Ref. & radius   & $E(B-V)$ & distance & log(age /yr) & [Fe/H] & pmra & pmdec & RV  \\
     & (arcmin) & (mag)  & (pc) 	  &               & (dex) & (mas/yr) & (mas/yr) & (km/s) \\\hline
\multicolumn{9}{c}{ASCC-92}\\\hline
1    & 18       & 0.25	 &	650	      &	9.01          &	      & -7.68$\pm$0.49  & -4.91$\pm$0.59 & \\
2	 & 13.4     & 0.312  &  644 	  &	9.08	      &       & -6.63$\pm$0.33  & -5,53$\pm$0.33 & \\	
3	 & 22.7 &	     & 678$\pm$20 &	8.54$\pm$0.02 & 0.25  &-1.157$\pm$0.306 & -4.115$\pm$0.280 & \\\hline
\multicolumn{9}{c}{Alessi~31}\\\hline
4    &          &       &             &               &       & -5.29$\pm$3.78 	& -2.16$\pm$4.67   & \\
5    &          &       &             &               &       & -5.26$\pm$5.45  & -2.20$\pm$6.52   & \\
6,7,8    &          &     0.56       &663.9$\pm$1.5&    8.38             &       & -1.164$\pm$0.225& -4.122$\pm$0.206 & 5.36$\pm$1.95 \\\hline
9    & {\bf }   & {\bf 0.45-0.92} & {\bf 676.0$\pm$10.0}      & {\bf 8.85$\pm$0.05}         & {\bf 0.00$\pm$0.10} &    {\bf -1.153$\pm$0.128} & {\bf -4.201$\pm$0.115}        & 
{\bf 0.98$\pm$9.81}\\\hline
\end{tabular}

\noindent Ref.: (1) \citet{kharchenkoetal2005};  (2) \citet{kharchenkoetal2013}; (3) \citet{lx2019};
(4) \citet{diasetal2014}; (5) \citet{sampedroetal2017}; (6) \citet{cantatgaudinetal2018}; 
(7) \citet{soubiranetal2018}; (8) \citet{cantatgaudinetal2020}; (9) this work.

\end{table*}

\begin{figure*}
\includegraphics[width=\textwidth]{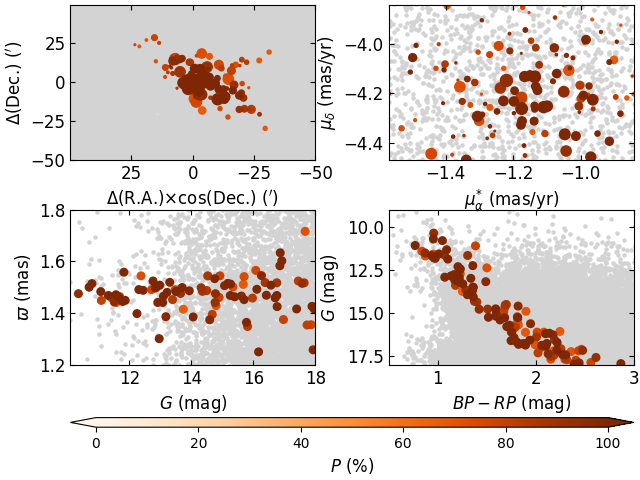}
\caption{Relationships of astrometric and photometric parameters of stars
in the field of ASCC~92. Points are from {\it Gaia} DR3, and those coloured
according to the cluster membership probability $P$ colour scale, 
are stars selected by HDBSCAN. Coloured symbols' size in top panels is
proportional to the star brightness.}
\label{fig1}
\end{figure*}

\section{Data analysis}

Machine learning techniques have become a powerful tool for discovering new open clusters
using large databases \citep{hr2021,jaehnigetal2021,haoetal2022,castroginardetal2022}. 
We tried to recognize star members of ASCC~92 using the
widely recommended HDBSCAN 
\citep[Hierarchical Density-Based Spatial Clustering of Applications with Noise,][]{campelloetal2013}
Gaussian mixture model technique \citep{hr2021}. Cluster stars were recovered after extensively 
trying with different groups of searching variables on the {\it Gaia} DR3
\citep{gaiaetal2016,gaiaetal2022b} database. Indeed, we firstly used only positions 
(RA, Dec.) as searching variables, and did not have success in identifying the bulk of the
clusters members. With the aim of improving the clustering search, we then included proper
motions, and the results did not improve the previous performance.
We then constrained the search to
stars located inside a circle of 50$\arcmin$ centred on the cluster, $G <$ 18 mag,  
and proper motion centred on (pmra,pmdec) = (-1.18,-4.15) mas/yr and within a circle of 0.4 
mas/yr.  Having the cluster area and, particularly, the proper motions constrained for a 
successful clustering search, HDBSCAN  carried out a satisfactory search.
The outcome of the position-proper motion driven search is shown in Fig.~\ref{fig1}. 
As can be seen, the resulting cluster 
colour-magnitude diagram (CMD) is affected by differential reddening. 

Recently, \citet{cantatgaudinetal2018} recovered the largest number of
cluster members identified up to date, 185, using the Unsupervised Photometric Membership Assignment 
in Stellar Clusters \citep[UPMASK][]{kmm2014} on {\it Gaia} DR2 data.
We provide a comparison analysis between their results and the present ones in the Appendix.
However, we refer the reader to the study 
of \citet{hr2021}, who carried out a sound comparison of the performance of different 
clustering searching algorithms, and 
described advantages and disadvantages of them.

\begin{figure}
\includegraphics[width=\columnwidth]{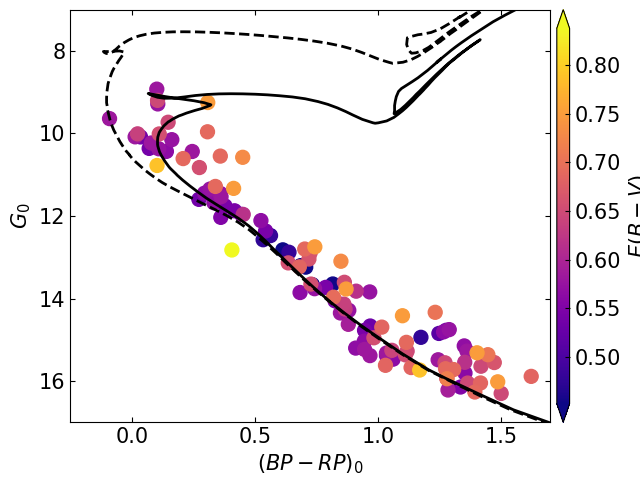}
\caption{ CMD of stars with cluster membership probabilities $P>$ 70 percent in the
field of ASCC~92 according to HDBSCAN, coloured according to the individual $E(B-V)$
reddening values. Solid and dashed lines correspond to
theoretical isochrones with a solar metallicity and log($t$ /yr) = 8.85 and 8.45, respectively.}
\label{fig2}
\end{figure}

We used different Milky Way reddening map models through the 
GALExtin\footnote{http://www.galextin.org/} interface
 \citep{amoresetal2021} to obtain the mean interstellar reddening
along the line of sight of the cluster, and surprisingly realised of the wide range of
mean $E(B-V)$ values retrieved, namely, 1.09 mag \citep{schlegeletal1998}; 0.28 mag
\citep{drimmeletal2003}; 0.15-1.83 mag \citep{al2005}; 4.9 mag \citep{marshalletal2006};
1.05 mag \citep{abergeletal2014}; 1.11 mag \citep{schlaflyetal2014}; and 0.67 mag 
\citep{chenetal2019}, respectively. Since the reddening maps of \citet{chenetal2019}
were built specifically for {\it Gaia} bandpasses, and have better spatial
resolution 6 arcmin than other reddening maps, we adopted them for interpolating individual reddening 
values for the selected cluster stars with membership probabilities $P >$ 70 percent and 
corrected their $G$ magnitudes and $BP-RP$  colours using the total to selective absorption 
ratios given by \citet{chenetal2019}. The absorption uncertainties ($\sigma$($A_G$) span
from 0.003 up to 0.020 mag, with an average of 0.010 mag at any $A_G$ interval.)
The reddening corrected cluster CMD is shown
in Fig.~\ref{fig2}, with stars coloured according to their $E(B-V)$ values. Note
that the spatial resolution of the reddening map allowed the detection of differential
reddening (see Fig.~\ref{fig1}).

In order to derive the cluster age and metallicity we fitted theoretical isochrones  
computed by \citet[][PARSEC v1.2S\footnote{http://stev.oapd.inaf.it/cgi-bin/cmd}]{betal12} 
and the Automated Stellar Cluster Analysis code \citep[\texttt{ASteCA,}][]{pvp15}. which
explored the parameter space  of 
synthetic CMDs through the minimization of the likelihood function defined by 
\citet[][the Poisson likelihood ratio (eq. 10)]{tremmeletal2013} using a parallel tempering 
Bayesian MCMC algorithm, and the optimal binning \citet{knuth2018}'s method.
Uncertainties in the resulting age and metallicity were estimated from the standard bootstrap 
method described in \citet{efron1982}.  While running \texttt{ASteCA}, we used
the mean cluster distance obtained from the cluster member stars' parallaxes.
ASCC~92 turned out to be a Hyades-like aged open cluster with a solar metal content.
Table~\ref{tab1} lists all the derived parameters, while Fig.~\ref{fig2} illustrates the matching 
performance of an isochrone with the derived age and metallicity of ASCC~92.
Table~\ref{tab1} aslo summarizes the different 
works carried out on the cluster under different cluster names, for comparison
purposes.

\begin{figure*}
\includegraphics[width=\columnwidth]{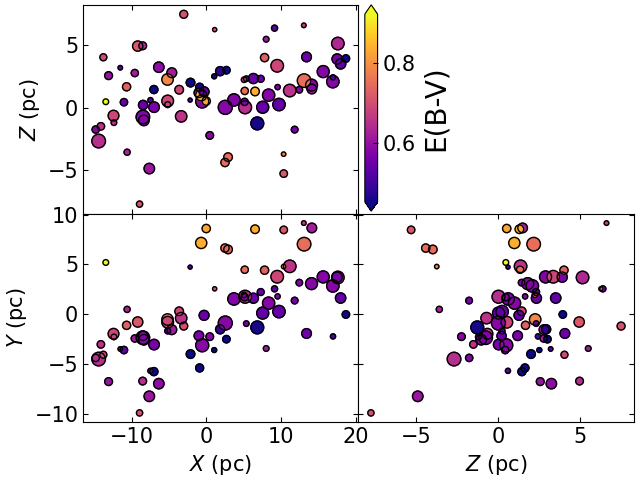}
\includegraphics[width=\columnwidth]{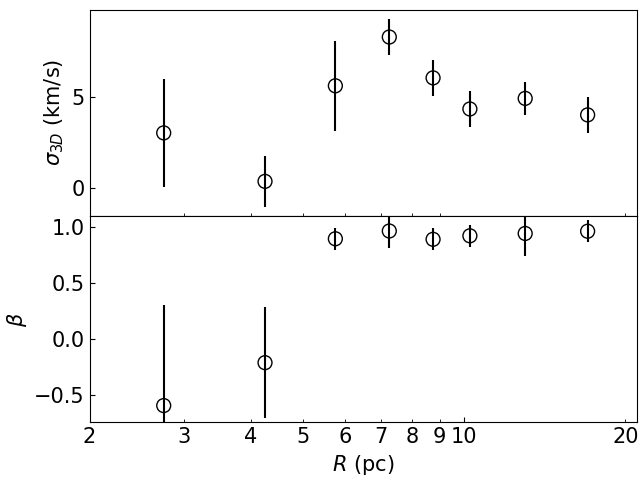}
\caption{{\it Left panel:} Spatial distribution of cluster members in the Galactic ($X,Y,Z$) framework
from the perspective of the cluster centre.
Symbol size is proportional to the star brightness, colored according to the
individual $E(B-V)$ values. {\it Right panel:} Derived 3D velocity dispersion (top) and anisotropy 
(bottom) for cluster members as a function of the distance to the cluster centre.}
\label{fig3}
\end{figure*}

From  {\it Gaia} DR3  coordinates, proper motions, parallaxes, and radial
velocities of cluster members, we computed
Galactic coordinates $(X,Y,Z)$ and space velocities $(V_X,V_Y,V_Z)$
employing the \texttt{Astropy}\footnote{http://www.astropy.org} package 
\citep{astropy2013,astropy2018}, which simply required the input of Right Ascension, 
Declination, parallaxes, proper motions and radial velocity of each star. We adopted the default
values for the Galactocentric coordinate frame, namely : ICRS coordinates (RA, DEC) of the
Galactic centre = (266.4051$\degr$, -28.936175$\degr$); Galactocentric distance of the Sun =
8.122 kpc, height of the Sun above the Galactic midplane = 20.8 pc; and solar motion relative to the
Galactic centre = (12.9, 245.6, 7.78) km/s. The position of the Sun is assumed to be on the $X$ 
axis of the final, right-handed system. That is, the $X$ axis points from the position of the Sun
projected to the Galactic midplane to the Galactic centre - roughly towards (l,b)=
(0$\degr$,0$\degr$). The $Y$ axis points roughly towards Galactic longitude {\it l}=90$\degr$, 
and the $Z$ axis points roughly towards the North Galactic Pole (b=90$\degr$).
\citet{gaiaetal2022c} extensively analysed the uncertainties in the Galactic coordinates as
a function of the heliocentric distance for stars with radial velocity measurements and found
that at the ASCC~92 distance the relative errors are less than 1.5 per cent.
The spatial distribution of the cluster
stars is depicted in Fig.~\ref{fig3}, where they were on purpose coloured according
 to their individual $E(B-V)$ values, aiming at rebuilding the 3D reddening map.

As for the space velocity components, they were transformed to the 
$V_\phi$, $V_\theta$, and $V_r$ spherical components, and from them we computed the 3D dispersion
velocity and anisotropy $\beta$, following the prescription described in
\citet{piatti2019}.  We obtained eight  3D dispersion velocity and
anisotropy profiles, respectively, using distance intervals of 1.0, 1.5, 2.0, 2.5, 3.0, 3.5, 
4.0 and 4.5 pc, where we calculated the values of these parameters. Then,
we obtained the mean values and standard dispersions (error bars) as a function of the
distance to the cluster centre ($R$) depicted in Fig.~\ref{fig3}.

\section{Results and Discussion}

\subsection{Cluster's fundamental parameters}

Thanks to the stringent {\it Gaia} DR3 data sets selection criteria applied 
\citep[see a comprehensive overview in][]{hr2021},
we identified by visual inspection in the vector point diagram cluster stars tightly gathered
around pmra and pmdec values that resulted to be fair first guesses of the derived mean 
cluster proper motion, in excellent agreement with recent accurate estimates
\citep{cantatgaudinetal2018,soubiranetal2018,cantatgaudinetal2020}. Because
of the highly variable reddening across the cluster field the combination of proper
motions and parallaxes resulted to be a key strategic search procedure, that allowed
us to recover the spatial distribution of  cluster's stars. Whenever sky positions were
used as HDBSCAN searching variables, we did not recover any clustering signature
in the phase-space used. This means that genuine clustering detection is not always 
guaranteed by enlarging the number of searching variables (e.g. position, motion, 
photometry).

Previous works on ASCC~92 did not report the existence of variable interstellar
absorption along the cluster line of sight, which we found it spans the $E(B-V)$ range from 
0.45 mag to 0.92 mag.
We speculate with the possibility that previous age cluster estimates based on
theoretical isochrone fit to the cluster CMD obtained much younger ages because
of this effect. We superimposed, for comparison purposes,
an isochrone of log($t$ /yr) = 8.45 (the average age estimate from \citet{lx2019}
and \citet{cantatgaudinetal2020}), which confirms that ASCC~92 is not such
a younger cluster. That younger isochrone seems to better reproduce the
observed cluster CMD of Fig.~\ref{fig1}. Likewise, cluster ages larger than 1 Gyr
(see Table~\ref{tab1}) could be affected by field star contamination.

\subsection{Cluster's motion}

The mean cluster space velocity turned out to be ($V_X$, $V_Y$, $V_Z$) =
(16.8, 232.9, 9.4) km/s, which shows that ASCC~92 is mainly moving following the
rotation of the Milky Way. If we drew the star velocity vectors in Fig.~\ref{fig3}, we
would see that the respective arrows are nearly pointing parallel to the Y axis.
According to the space distribution of $E(B-V)$ values, it would seem that ASCC~92
is moving toward a direction of increasing interstellar absorption.
From Fig.~\ref{fig2}, we estimated that $E(B-V)$ varies $\sim$ 0.2 mag in 5 pc. By using 
$E(B-V)$ = $N$/(5.8$\times$10$^{21}$  atoms/cm$^2$ per mag) \citep{bohlinetal1978}, we
got an HI column density of 1.16$\times$10$^{21}$ atoms/cm$^2$, which along the 5 pc
gives a space density of 75 atoms/cm$^{3}$. This is a diffuse cloud density, or a density 
for the edge of a molecular cloud. The juxtaposition of the cluster next to a moderate density gas 
cloud might be expected if the cluster were young (age $<$ 100 Myr), since the gas could be from 
the GMC where it formed. \citet[][and references therein]{chenetal2020} compiled a
large catalogue of molecular clouds with accurate distances within 4 kpc of the
Milky Way disc. We did not find any molecular cloud that matches the ASCC~92 locus, and
did not find any expected motion signatures in the distribution of the space velocity vectors
that accounts for a collision of the cluster with a molecular cloud. Therefore, a
plausible interpretation could be that ASCC~92 is entering a more obscure region of 
the Milky Way disc.

\subsection{Cluster's internal kinematic}

We analysed the cluster internal kinematics and found that there is no signature of expansion 
along its major axis, which is contained mainly in the ($X,Y$) plane, nor any expected pattern driven by
Milky Way tidal forces \citep{meingastetal2021}. As can be seen in Fig.~\ref{fig3} (right panel),
the cluster's stars motion is isotropic from $\sim$ 6 pc outwards from the cluster centre, which
means that there is no privileged direction of motion. Cluster stars placed in the inner 
$\sim$ 5 pc 
show the expected decreasing velocity dispersion profile of bound stars, while the velocity 
dispersion increases outwards. Such a peculiar behaviour was predicted by \citet{kupperetal2010b}
and, as far as we are aware, it has not been observed until now. 

\begin{figure}
\includegraphics[width=\columnwidth]{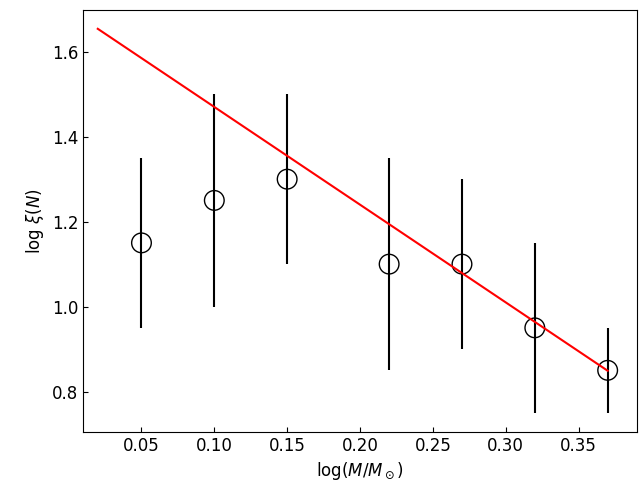}
\caption{Cluster mass function built using stars with membership probabilities $>$ 70 per cent
(see text for details). The red line represents the fitted \citet{kroupa02}'s mass function.}
\label{fig4}
\end{figure}

With the aim of verifying such a
theoretical result, we computed the critical energy a star needs to escape the cluster and its
total energy. For the calculation we used equations 13 and 14 of \citet{kupperetal2010b}, 
adopting a Milky Way mass enclosed inside the cluster Galactocentric distance (7.66 kpc) of 
(1.0$\pm$0.2)$\times$10$^{11}$$\msun$ \citep{birdetal2022}. An upper cluster mass was
calculated from the cluster mass function built from stars with membership probabilities 
$>$ 70 per cent. We obtained the individual masses by interpolation in the adopted PARSEC  
v1.2S isochrone.  We then fitted the resulting mass function with a \citet{kroupa02}'s profile. 
Fig.~\ref{fig4}
shows the resulting mass function, with points and error bars representing the average and
corresponding standard deviations of five mass functions built using mass intervals of 
log($M$/\msun) = 0.05 up to 0.30, in steps of 0.05. We finally integrated the fitted 
\citet{kroupa02}'s mass function from the maximum observed mass down to 0.5$\msun$.
The upper cluster mass turned out to be 580 $\msun$. We used alternatively \texttt{ASteCA}
(see Sect. 2) for the construction of a large number of synthetic CMDs from which it finds 
the one which best resembles the observed CMD. Thus, the star 
cluster present mass and the binary fraction associated to that best representative
generated synthetic CMD are  adopted as the best-fitted star cluster properties. 
\texttt{ASteCA} generates synthetic CMDs by adopting the initial mass function given by 
\citet{kroupa02} and a minimum mass ratio for the generation of binaries of 0.5. The total
observed star cluster mass and its binary fraction were set in the ranges 100-5000 $\msun$ 
and 0.0-0.5, respectively. We derived a cluster mass of 566$\pm$110$\msun$ with a binary fraction
of 0.40$\pm$0.06, in excellent agreement with the above mass estimate.

\begin{figure}
\includegraphics[width=\columnwidth]{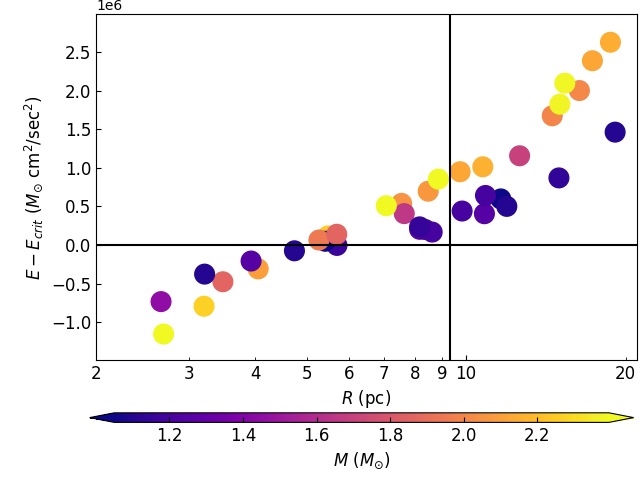}
\caption{Relationship between the difference of total to critical star energy versus
distance to the cluster centre. The vertical line represent the $r_J$. Symbols are
coloured according to the star masses.}
\label{fig5}
\end{figure}

Fig.~\ref{fig5} shows the difference between the total and the critical star energy as a function of
the distance to the cluster centre. The vertical line represents the derived $r_J$ value
(9.3 pc). As can be seen, stars located at $R <$ 5 pc are bound stars, in excellent agreement with
the 3D velocity dispersion profile and calculated cluster anisotropy. Stars beyond
$r_J$ are unbound stars, which show a hot kinematics, while those located at
5 $<$ $R$ (pc) $<$ 9.3 are the stars that will next escape the cluster. We added the
masses of stars enclosed
within $r_J$,  which resulted to be 56$\msun$, nearly 10 percent of the upper cluster mass limit. This outcome
suggests that ASCC~92 is facing its last breath.


\section{Conclusions}

We used {\it Gaia} DR3 data to thoroughly analyse ASCC~92, a Milky Way open
cluster with somewhat discrepant parameters in the literature. The analysis of the
{\it Gaia} data resulted challenging while trying to employ the recommended 
Hierarchical Density-Based Spatial Clustering of Applications with Noise, because
the cluster is projected toward a region affected by noticeable differential reddening.
Such a variable obscure appearance of ASCC~92 caused that clustering in the
stellar distribution in the sky is not evident. Therefore, by using positions
as clustering searching variables, we arrived to an unsatisfactory number of identification 
of cluster members. By imposing some constraints, namely, a limited region of the 
searchable area, mean proper motions and dispersion in very good agreement with 
reliable estimates, we identified 171 stars with similar parallaxes that clearly 
stand out in the sky as an stellar aggregate. Differential reddening has not been used 
in previous analyses of the
cluster's CMD. This made the evolved region of the cluster Main Sequence,
and particularly, the Main Sequence turnoff,  appear blurred. With the appropriate
correction for differential reddening, we obtained a cluster age 710 Myr older than
recent estimates using {\it Gaia} data 240 Myr. The isochrone corresponding to the cluster
age matches satisfactorily well the cluster Main Sequence along $\sim$ 7 mag.

 We also studied the cluster internal dynamical evolution and found that ASCC~92, while
moving nearly in the Milky Way disc following Galactic rotation, does not show
any signature of internal expansion, rotation, or both combined. Its 3D velocity
dispersion profile and anisotropy reveal that innermost cluster members move following
the expected kinematics of bound stars, while those in the cluster outskirts show
an isotropic kinematic behaviour, like unbound stars. ASCC~92 shows a kinematic transition
region of increasing velocity dispersion and anisotropy of bound stars. These stars
have the necessary energy to leave the cluster, but they are still within its boundaries.
This peculiarity was predicted from N-body simulations by \citet{kupperetal2010b}.
As fas as we are aware, ASCC~92 is the first studied clusters showing evidence of
being populated by these kinematically peculiar stars.

\section*{Acknowledgements}
I thank Enrico Vesperini, Bruce Elmegreen and Pavel Kroupa for insightful comments
and suggestions.
I thank the referee for the thorough reading of the manuscript and
timely suggestions to improve it. 

This work has made use of data from the European Space Agency (ESA) mission
{\it Gaia} (\url{https://www.cosmos.esa.int/gaia}), processed by the {\it Gaia}
Data Processing and Analysis Consortium (DPAC,
\url{https://www.cosmos.esa.int/web/gaia/dpac/consortium}). Funding for the DPAC
has been provided by national institutions, in particular the institutions
participating in the {\it Gaia} Multilateral Agreement.

This research made use of Astropy, a community-developed
core Python package for Astronomy.

\section{Data availability}

Data used in this work are available upon request to the author.




\begin{thebibliography}{}
\makeatletter
\relax
\def\mn@urlcharsother{\let\do\@makeother \do\$\do\&\do\#\do\^\do\_\do\%\do\~}
\def\mn@doi{\begingroup\mn@urlcharsother \@ifnextchar [ {\mn@doi@}
  {\mn@doi@[]}}
\def\mn@doi@[#1]#2{\def\@tempa{#1}\ifx\@tempa\@empty \href
  {http://dx.doi.org/#2} {doi:#2}\else \href {http://dx.doi.org/#2} {#1}\fi
  \endgroup}
\def\mn@eprint#1#2{\mn@eprint@#1:#2::\@nil}
\def\mn@eprint@arXiv#1{\href {http://arxiv.org/abs/#1} {{\tt arXiv:#1}}}
\def\mn@eprint@dblp#1{\href {http://dblp.uni-trier.de/rec/bibtex/#1.xml}
  {dblp:#1}}
\def\mn@eprint@#1:#2:#3:#4\@nil{\def\@tempa {#1}\def\@tempb {#2}\def\@tempc
  {#3}\ifx \@tempc \@empty \let \@tempc \@tempb \let \@tempb \@tempa \fi \ifx
  \@tempb \@empty \def\@tempb {arXiv}\fi \@ifundefined
  {mn@eprint@\@tempb}{\@tempb:\@tempc}{\expandafter \expandafter \csname
  mn@eprint@\@tempb\endcsname \expandafter{\@tempc}}}

\bibitem[\protect\citeauthoryear{{Alessi}, {Moitinho}  \& {Dias}}{{Alessi}
  et~al.}{2003}]{alessietal2003}
{Alessi} B.~S.,  {Moitinho} A.,   {Dias} W.~S.,  2003, \mn@doi [\aap]
  {10.1051/0004-6361:20031280}, \href
  {https://ui.adsabs.harvard.edu/abs/2003A&A...410..565A} {410, 565}

\bibitem[\protect\citeauthoryear{{Am{\^o}res} \& {L{\'e}pine}}{{Am{\^o}res} \&
  {L{\'e}pine}}{2005}]{al2005}
{Am{\^o}res} E.~B.,  {L{\'e}pine} J.~R.~D.,  2005, \mn@doi [\aj]
  {10.1086/430957}, \href
  {https://ui.adsabs.harvard.edu/abs/2005AJ....130..659A} {130, 659}

\bibitem[\protect\citeauthoryear{{Am{\^o}res} et~al.,}{{Am{\^o}res}
  et~al.}{2021}]{amoresetal2021}
{Am{\^o}res} E.~B.,  et~al., 2021, \mn@doi [\mnras] {10.1093/mnras/stab2248},
  \href {https://ui.adsabs.harvard.edu/abs/2021MNRAS.508.1788A} {508, 1788}

\bibitem[\protect\citeauthoryear{{Astropy Collaboration} et~al.,}{{Astropy
  Collaboration} et~al.}{2013}]{astropy2013}
{Astropy Collaboration} et~al., 2013, \mn@doi [\aap]
  {10.1051/0004-6361/201322068}, \href
  {https://ui.adsabs.harvard.edu/abs/2013A&A...558A..33A} {558, A33}

\bibitem[\protect\citeauthoryear{{Astropy Collaboration} et~al.,}{{Astropy
  Collaboration} et~al.}{2018}]{astropy2018}
{Astropy Collaboration} et~al., 2018, \mn@doi [\aj] {10.3847/1538-3881/aabc4f},
  \href {https://ui.adsabs.harvard.edu/abs/2018AJ....156..123A} {156, 123}

\bibitem[\protect\citeauthoryear{{Babusiaux} et~al.,}{{Babusiaux}
  et~al.}{2022}]{gaiaetal2022b}
{Babusiaux} C.,  et~al., 2022, arXiv e-prints, \href
  {https://ui.adsabs.harvard.edu/abs/2022arXiv220605989B} {p. arXiv:2206.05989}

\bibitem[\protect\citeauthoryear{{Bird} et~al.,}{{Bird}
  et~al.}{2022}]{birdetal2022}
{Bird} S.~A.,  et~al., 2022, \mn@doi [\mnras] {10.1093/mnras/stac2036}, \href
  {https://ui.adsabs.harvard.edu/abs/2022MNRAS.tmp.1951B} {}

\bibitem[\protect\citeauthoryear{{Bohlin}, {Savage}  \& {Drake}}{{Bohlin}
  et~al.}{1978}]{bohlinetal1978}
{Bohlin} R.~C.,  {Savage} B.~D.,   {Drake} J.~F.,  1978, \mn@doi [\apj]
  {10.1086/156357}, \href
  {https://ui.adsabs.harvard.edu/abs/1978ApJ...224..132B} {224, 132}

\bibitem[\protect\citeauthoryear{{Bovy}}{{Bovy}}{2015}]{bovy2015}
{Bovy} J.,  2015, \mn@doi [\apjs] {10.1088/0067-0049/216/2/29}, \href
  {https://ui.adsabs.harvard.edu/abs/2015ApJS..216...29B} {216, 29}

\bibitem[\protect\citeauthoryear{{Bressan}, {Marigo}, {Girardi}, {Salasnich},
  {Dal Cero}, {Rubele}  \& {Nanni}}{{Bressan} et~al.}{2012}]{betal12}
{Bressan} A.,  {Marigo} P.,  {Girardi} L.,  {Salasnich} B.,  {Dal Cero} C.,
  {Rubele} S.,   {Nanni} A.,  2012, \mn@doi [\mnras]
  {10.1111/j.1365-2966.2012.21948.x}, 427, 127

\bibitem[\protect\citeauthoryear{{Campello}, {Moulavi}  \& {Sander}}{{Campello}
  et~al.}{2013}]{campelloetal2013}
{Campello} R.~J.~G.~B.,  {Moulavi} D.,   {Sander} J., 2013, \reference

\bibitem[\protect\citeauthoryear{{Cantat-Gaudin} et~al.,}{{Cantat-Gaudin}
  et~al.}{2018}]{cantatgaudinetal2018}
{Cantat-Gaudin} T.,  et~al., 2018, \mn@doi [\aap]
  {10.1051/0004-6361/201833476}, \href
  {https://ui.adsabs.harvard.edu/abs/2018A&A...618A..93C} {618, A93}

\bibitem[\protect\citeauthoryear{{Cantat-Gaudin} et~al.,}{{Cantat-Gaudin}
  et~al.}{2020}]{cantatgaudinetal2020}
{Cantat-Gaudin} T.,  et~al., 2020, \mn@doi [\aap]
  {10.1051/0004-6361/202038192}, \href
  {https://ui.adsabs.harvard.edu/abs/2020A&A...640A...1C} {640, A1}

\bibitem[\protect\citeauthoryear{{Castro-Ginard} et~al.,}{{Castro-Ginard}
  et~al.}{2022}]{castroginardetal2022}
{Castro-Ginard} A.,  et~al., 2022, \mn@doi [\aap]
  {10.1051/0004-6361/202142568}, \href
  {https://ui.adsabs.harvard.edu/abs/2022A&A...661A.118C} {661, A118}

\bibitem[\protect\citeauthoryear{{Chen} et~al.,}{{Chen}
  et~al.}{2019}]{chenetal2019}
{Chen} B.~Q.,  et~al., 2019, \mn@doi [\mnras] {10.1093/mnras/sty3341}, \href
  {https://ui.adsabs.harvard.edu/abs/2019MNRAS.483.4277C} {483, 4277}

\bibitem[\protect\citeauthoryear{{Chen} et~al.,}{{Chen}
  et~al.}{2020}]{chenetal2020}
{Chen} B.~Q.,  et~al., 2020, \mn@doi [\mnras] {10.1093/mnras/staa235}, \href
  {https://ui.adsabs.harvard.edu/abs/2020MNRAS.493..351C} {493, 351}

\bibitem[\protect\citeauthoryear{{Dias}, {Monteiro}, {Caetano}, {L{\'e}pine},
  {Assafin}  \& {Oliveira}}{{Dias} et~al.}{2014}]{diasetal2014}
{Dias} W.~S.,  {Monteiro} H.,  {Caetano} T.~C.,  {L{\'e}pine} J.~R.~D.,
  {Assafin} M.,   {Oliveira} A.~F.,  2014, \mn@doi [\aap]
  {10.1051/0004-6361/201323226}, \href
  {https://ui.adsabs.harvard.edu/abs/2014A&A...564A..79D} {564, A79}

\bibitem[\protect\citeauthoryear{{Drimmel}, {Cabrera-Lavers}  \&
  {L{\'o}pez-Corredoira}}{{Drimmel} et~al.}{2003}]{drimmeletal2003}
{Drimmel} R.,  {Cabrera-Lavers} A.,   {L{\'o}pez-Corredoira} M.,  2003, \mn@doi
  [\aap] {10.1051/0004-6361:20031070}, \href
  {https://ui.adsabs.harvard.edu/abs/2003A&A...409..205D} {409, 205}

\bibitem[\protect\citeauthoryear{{Efron}}{{Efron}}{1982}]{efron1982}
{Efron} B.,  1982, {The Jackknife, the Bootstrap and other resampling plans}

\bibitem[\protect\citeauthoryear{{Gaia Collaboration} et~al.,}{{Gaia
  Collaboration} et~al.}{2016}]{gaiaetal2016}
{Gaia Collaboration} et~al., 2016, \mn@doi [\aap]
  {10.1051/0004-6361/201629272}, \href
  {http://adsabs.harvard.edu/abs/2016A%26A...595A...1G} {595, A1}

\bibitem[\protect\citeauthoryear{{Gaia Collaboration} et~al.,}{{Gaia
  Collaboration} et~al.}{2022}]{gaiaetal2022c}
{Gaia Collaboration} et~al., 2022, arXiv e-prints, \href
  {https://ui.adsabs.harvard.edu/abs/2022arXiv220606207G} {p. arXiv:2206.06207}

\bibitem[\protect\citeauthoryear{{Hao}, {Xu}, {Wu}, {Lin}, {Liu}  \&
  {Li}}{{Hao} et~al.}{2022}]{haoetal2022}
{Hao} C.~J.,  {Xu} Y.,  {Wu} Z.~Y.,  {Lin} Z.~H.,  {Liu} D.~J.,   {Li} Y.~J.,
  2022, \mn@doi [\aap] {10.1051/0004-6361/202243091}, \href
  {https://ui.adsabs.harvard.edu/abs/2022A&A...660A...4H} {660, A4}

\bibitem[\protect\citeauthoryear{{H{\o}g}}{{H{\o}g}}{2000}]{hog2000}
{H{\o}g} E.,  2000, in {Murdin} P.,  ed., , Encyclopedia of Astronomy and
  Astrophysics.
p.~2862, \mn@doi{10.1888/0333750888/2862}

\bibitem[\protect\citeauthoryear{{Hunt} \& {Reffert}}{{Hunt} \&
  {Reffert}}{2021}]{hr2021}
{Hunt} E.~L.,  {Reffert} S.,  2021, \mn@doi [\aap]
  {10.1051/0004-6361/202039341}, \href
  {https://ui.adsabs.harvard.edu/abs/2021A&A...646A.104H} {646, A104}

\bibitem[\protect\citeauthoryear{{Jaehnig}, {Bird}  \&
  {Holley-Bockelmann}}{{Jaehnig} et~al.}{2021}]{jaehnigetal2021}
{Jaehnig} K.,  {Bird} J.,   {Holley-Bockelmann} K.,  2021, \mn@doi [\apj]
  {10.3847/1538-4357/ac1d51}, \href
  {https://ui.adsabs.harvard.edu/abs/2021ApJ...923..129J} {923, 129}

\bibitem[\protect\citeauthoryear{{Jerabkova}, {Boffin}, {Beccari}, {de Marchi},
  {de Bruijne}  \& {Prusti}}{{Jerabkova} et~al.}{2021}]{jerabkovaetal2021}
{Jerabkova} T.,  {Boffin} H. M.~J.,  {Beccari} G.,  {de Marchi} G.,  {de
  Bruijne} J. H.~J.,   {Prusti} T.,  2021, \mn@doi [\aap]
  {10.1051/0004-6361/202039949}, \href
  {https://ui.adsabs.harvard.edu/abs/2021A&A...647A.137J} {647, A137}

\bibitem[\protect\citeauthoryear{{Kharchenko}}{{Kharchenko}}{2001}]{k01}
{Kharchenko} N.~V.,  2001, Kinematika i Fizika Nebesnykh Tel, \href
  {https://ui.adsabs.harvard.edu/abs/2001KFNT...17..409K} {17, 409}

\bibitem[\protect\citeauthoryear{{Kharchenko}, {Piskunov}, {R{\"o}ser},
  {Schilbach}  \& {Scholz}}{{Kharchenko} et~al.}{2005}]{kharchenkoetal2005}
{Kharchenko} N.~V.,  {Piskunov} A.~E.,  {R{\"o}ser} S.,  {Schilbach} E.,
  {Scholz} R.~D.,  2005, \mn@doi [\aap] {10.1051/0004-6361:20052740}, \href
  {https://ui.adsabs.harvard.edu/abs/2005A&A...440..403K} {440, 403}

\bibitem[\protect\citeauthoryear{{Kharchenko}, {Berczik}, {Petrov}, {Piskunov},
  {R{\"o}ser}, {Schilbach}  \& {Scholz}}{{Kharchenko}
  et~al.}{2009}]{kharchenkoetal2009b}
{Kharchenko} N.~V.,  {Berczik} P.,  {Petrov} M.~I.,  {Piskunov} A.~E.,
  {R{\"o}ser} S.,  {Schilbach} E.,   {Scholz} R.~D.,  2009, \mn@doi [\aap]
  {10.1051/0004-6361/200810407}, \href
  {https://ui.adsabs.harvard.edu/abs/2009A&A...495..807K} {495, 807}

\bibitem[\protect\citeauthoryear{{Kharchenko}, {Piskunov}, {Schilbach},
  {R{\"o}ser}  \& {Scholz}}{{Kharchenko} et~al.}{2012}]{kharchenkoetal2012}
{Kharchenko} N.~V.,  {Piskunov} A.~E.,  {Schilbach} E.,  {R{\"o}ser} S.,
  {Scholz} R.~D.,  2012, \mn@doi [\aap] {10.1051/0004-6361/201118708}, \href
  {https://ui.adsabs.harvard.edu/abs/2012A&A...543A.156K} {543, A156}

\bibitem[\protect\citeauthoryear{{Kharchenko}, {Piskunov}, {Schilbach},
  {R{\"o}ser}  \& {Scholz}}{{Kharchenko} et~al.}{2013}]{kharchenkoetal2013}
{Kharchenko} N.~V.,  {Piskunov} A.~E.,  {Schilbach} E.,  {R{\"o}ser} S.,
  {Scholz} R.~D.,  2013, \mn@doi [\aap] {10.1051/0004-6361/201322302}, \href
  {https://ui.adsabs.harvard.edu/abs/2013A&A...558A..53K} {558, A53}

\bibitem[\protect\citeauthoryear{{Knuth}}{{Knuth}}{2018}]{knuth2018}
{Knuth} K.~H.,  2018, {optBINS: Optimal Binning for histograms} (\mn@eprint
  {ascl} {1803.013})

\bibitem[\protect\citeauthoryear{{Kronberger} et~al.,}{{Kronberger}
  et~al.}{2006}]{kronbergeretal2006}
{Kronberger} M.,  et~al., 2006, \mn@doi [\aap] {10.1051/0004-6361:20054057},
  \href {https://ui.adsabs.harvard.edu/abs/2006A&A...447..921K} {447, 921}

\bibitem[\protect\citeauthoryear{{Krone-Martins} \& {Moitinho}}{{Krone-Martins}
  \& {Moitinho}}{2014}]{kmm2014}
{Krone-Martins} A.,  {Moitinho} A.,  2014, \mn@doi [\aap]
  {10.1051/0004-6361/201321143}, \href
  {https://ui.adsabs.harvard.edu/abs/2014A&A...561A..57K} {561, A57}

\bibitem[\protect\citeauthoryear{Kroupa}{Kroupa}{2002}]{kroupa02}
Kroupa P.,  2002, \mn@doi [Science] {10.1126/science.1067524}, 295, 82

\bibitem[\protect\citeauthoryear{{K{\"u}pper}, {Kroupa}, {Baumgardt}  \&
  {Heggie}}{{K{\"u}pper} et~al.}{2010}]{kupperetal2010b}
{K{\"u}pper} A. H.~W.,  {Kroupa} P.,  {Baumgardt} H.,   {Heggie} D.~C.,  2010,
  \mn@doi [\mnras] {10.1111/j.1365-2966.2010.17084.x}, \href
  {https://ui.adsabs.harvard.edu/abs/2010MNRAS.407.2241K} {407, 2241}

\bibitem[\protect\citeauthoryear{{Liu} \& {Pang}}{{Liu} \&
  {Pang}}{2019}]{lx2019}
{Liu} L.,  {Pang} X.,  2019, \mn@doi [\apjs] {10.3847/1538-4365/ab530a}, \href
  {https://ui.adsabs.harvard.edu/abs/2019ApJS..245...32L} {245, 32}

\bibitem[\protect\citeauthoryear{{Marshall}, {Robin}, {Reyl{\'e}}, {Schultheis}
   \& {Picaud}}{{Marshall} et~al.}{2006}]{marshalletal2006}
{Marshall} D.~J.,  {Robin} A.~C.,  {Reyl{\'e}} C.,  {Schultheis} M.,   {Picaud}
  S.,  2006, \mn@doi [\aap] {10.1051/0004-6361:20053842}, \href
  {https://ui.adsabs.harvard.edu/abs/2006A&A...453..635M} {453, 635}

\bibitem[\protect\citeauthoryear{{Meingast}, {Alves}  \&
  {Rottensteiner}}{{Meingast} et~al.}{2021}]{meingastetal2021}
{Meingast} S.,  {Alves} J.,   {Rottensteiner} A.,  2021, \mn@doi [\aap]
  {10.1051/0004-6361/202038610}, \href
  {https://ui.adsabs.harvard.edu/abs/2021A&A...645A..84M} {645, A84}

\bibitem[\protect\citeauthoryear{{Perren}, {V{\'a}zquez}  \& {Piatti}}{{Perren}
  et~al.}{2015}]{pvp15}
{Perren} G.~I.,  {V{\'a}zquez} R.~A.,   {Piatti} A.~E.,  2015, \mn@doi [\aap]
  {10.1051/0004-6361/201424946}, \href
  {http://adsabs.harvard.edu/abs/2015A%26A...576A...6P} {576, A6}

\bibitem[\protect\citeauthoryear{{Piatti}}{{Piatti}}{2019}]{piatti2019}
{Piatti} A.~E.,  2019, \mn@doi [\apj] {10.3847/1538-4357/ab3574}, \href
  {https://ui.adsabs.harvard.edu/abs/2019ApJ...882...98P} {882, 98}

\bibitem[\protect\citeauthoryear{{Planck Collaboration} et~al.,}{{Planck
  Collaboration} et~al.}{2014}]{abergeletal2014}
{Planck Collaboration} et~al., 2014, \mn@doi [\aap]
  {10.1051/0004-6361/201323195}, \href
  {https://ui.adsabs.harvard.edu/abs/2014A&A...571A..11P} {571, A11}

\bibitem[\protect\citeauthoryear{{Portegies Zwart} \& {McMillan}}{{Portegies
  Zwart} \& {McMillan}}{2018}]{pzm2018}
{Portegies Zwart} S.,  {McMillan} S.,  2018, {Astrophysical Recipes; The art of
  AMUSE}, \mn@doi{10.1088/978-0-7503-1320-9.
}

\bibitem[\protect\citeauthoryear{{R{\"o}ser} \& {Schilbach}}{{R{\"o}ser} \&
  {Schilbach}}{2019a}]{roseretal2019}
{R{\"o}ser} S.,  {Schilbach} E.,  2019a, \mn@doi [\aap]
  {10.1051/0004-6361/201935502}, \href
  {https://ui.adsabs.harvard.edu/abs/2019A&A...627A...4R} {627, A4}

\bibitem[\protect\citeauthoryear{{R{\"o}ser} \& {Schilbach}}{{R{\"o}ser} \&
  {Schilbach}}{2019b}]{rs2019}
{R{\"o}ser} S.,  {Schilbach} E.,  2019b, \mn@doi [\aap]
  {10.1051/0004-6361/201935502}, \href
  {https://ui.adsabs.harvard.edu/abs/2019A&A...627A...4R} {627, A4}

\bibitem[\protect\citeauthoryear{{Sampedro}, {Dias}, {Alfaro}, {Monteiro}  \&
  {Molino}}{{Sampedro} et~al.}{2017}]{sampedroetal2017}
{Sampedro} L.,  {Dias} W.~S.,  {Alfaro} E.~J.,  {Monteiro} H.,   {Molino} A.,
  2017, \mn@doi [\mnras] {10.1093/mnras/stx1485}, \href
  {https://ui.adsabs.harvard.edu/abs/2017MNRAS.470.3937S} {470, 3937}

\bibitem[\protect\citeauthoryear{{Schlafly} et~al.,}{{Schlafly}
  et~al.}{2014}]{schlaflyetal2014}
{Schlafly} E.~F.,  et~al., 2014, \mn@doi [\apj] {10.1088/0004-637X/789/1/15},
  \href {https://ui.adsabs.harvard.edu/abs/2014ApJ...789...15S} {789, 15}

\bibitem[\protect\citeauthoryear{{Schlegel}, {Finkbeiner}  \&
  {Davis}}{{Schlegel} et~al.}{1998}]{schlegeletal1998}
{Schlegel} D.~J.,  {Finkbeiner} D.~P.,   {Davis} M.,  1998, \mn@doi [\apj]
  {10.1086/305772}, \href {http://adsabs.harvard.edu/abs/1998ApJ...500..525S}
  {500, 525}

\bibitem[\protect\citeauthoryear{{Soubiran} et~al.,}{{Soubiran}
  et~al.}{2018}]{soubiranetal2018}
{Soubiran} C.,  et~al., 2018, \mn@doi [\aap] {10.1051/0004-6361/201834020},
  \href {https://ui.adsabs.harvard.edu/abs/2018A&A...619A.155S} {619, A155}

\bibitem[\protect\citeauthoryear{{Tremmel} et~al.,}{{Tremmel}
  et~al.}{2013}]{tremmeletal2013}
{Tremmel} M.,  et~al., 2013, \mn@doi [\apj] {10.1088/0004-637X/766/1/19}, \href
  {https://ui.adsabs.harvard.edu/abs/2013ApJ...766...19T} {766, 19}

\bibitem[\protect\citeauthoryear{{Wang} \& {Jerabkova}}{{Wang} \&
  {Jerabkova}}{2021}]{wj2021}
{Wang} L.,  {Jerabkova} T.,  2021, \mn@doi [\aap]
  {10.1051/0004-6361/202141838}, \href
  {https://ui.adsabs.harvard.edu/abs/2021A&A...655A..71W} {655, A71}

\bibitem[\protect\citeauthoryear{{Wang}, {Iwasawa}, {Nitadori}  \&
  {Makino}}{{Wang} et~al.}{2020}]{wangetal2020a}
{Wang} L.,  {Iwasawa} M.,  {Nitadori} K.,   {Makino} J.,  2020, \mn@doi
  [\mnras] {10.1093/mnras/staa1915}, \href
  {https://ui.adsabs.harvard.edu/abs/2020MNRAS.497..536W} {497, 536}

\makeatother
\end{thebibliography}




\appendix

\section{Membership criteria}

\begin{figure}
\includegraphics[width=\columnwidth]{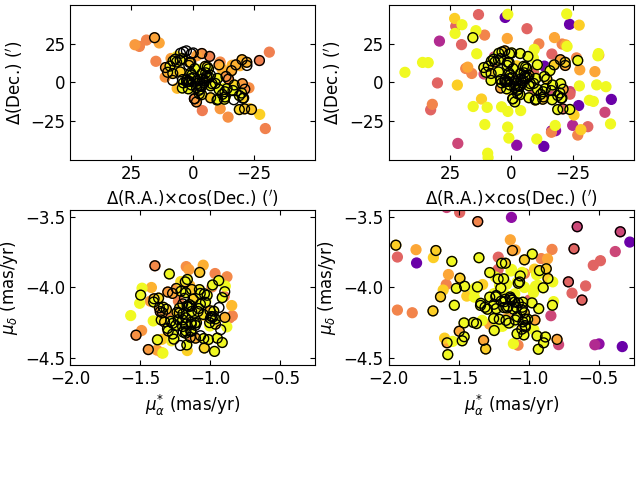}
\includegraphics[width=\columnwidth]{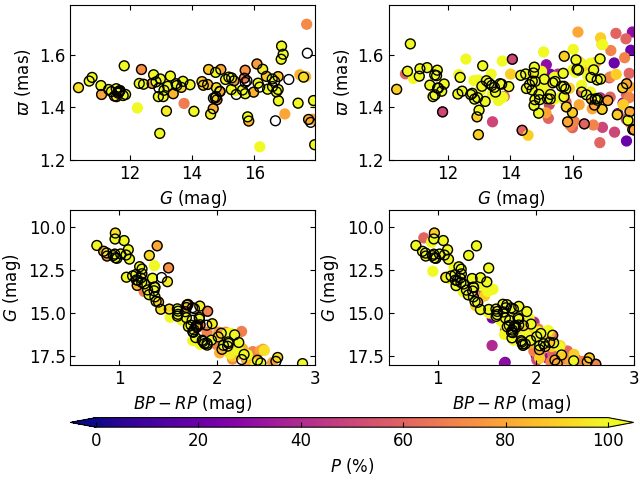}
\caption{ Relationships of {\it Gaia} parameters used in this work (left
panels) and in \citet{cantatgaudinetal2018} (right panels), respectively.}
\label{appendix}
\end{figure}

For completeness purposes, we compared our membership probabilities with those obtained by
\citet{cantatgaudinetal2018}. In order to cross-match both samples,  we employed the names
used by {\it Gaia} to identify stellar sources and the \texttt{Astropy} package.
Left and right panels of Fig.~\ref{appendix} correspond to the star selection made in this
work and those from \citet{cantatgaudinetal2018}, respectively. The symbols are coloured
according to the membership probability ($P$) assigned by each study, while those
encircled with a black open circle are the stars in common in both works. As can be seen, the
number of stars with relatively high membership probabilities is larger in the
 present sample.  The proper motions and mean parallaxes span 
wider ranges  in \citet{cantatgaudinetal2018}'s sample.  We are confident in 
the more stringent membership selection criteria applied here and the
use of stars with $P$ $>$ 70 per cent in the performed analyses.  Note that
the differential reddening is present in both CMDs. Finally, Table~\ref{tab2} lists
the {\it Gaia} names of the stars identified in this work that are absent in the
\citet{cantatgaudinetal2018}'s sample.

\begin{table*}
\caption{{\it Gaia} IDs of stars selected in this work that are absent in the
\citet{cantatgaudinetal2018}'s sample.}
\label{tab2}
\begin{tabular}{@{}cccc}\hline
4151598199606998528 &4151598440125176192 & 4151599677077307136& 4151581638213227264\\
4151573872910627584 &4151703855796778752 & 4151706535856460544& 4163590534247506560\\
4151550675772022528 &4150045173804666368 & 4151707704087604352& 4151559373100984704\\
4151569917226328448 &4151572051844491904 & 4150048884656569472& 4151584249552588416\\
4151693715358252288 &4151599230392430848 & 4151574285227815040& 4151561056728596480\\
4151598818083966464 &4163589851366486144 & 4151558342294713216& 4151584932425464832\\
4151607408016493568 &4151546075881760640 & 4163613628306249088& 4151548545465664000\\
4151594591833995776 &4151611874782646656 & 4151596614737584256& \\\hline

\end{tabular}
\end{table*}



\bsp	
\label{lastpage}
\end{document}